# Radar detection of wake vortex behind the aircraft: the detection range problem

Jiangkun Gong, Jun Yan*, Deyong Kong, and Deren Li

*Abstract*— In this study, we showcased the detection of the wake vortex produced by a medium aircraft at distances exceeding 10 km using an X-band pulse-Doppler radar. We analyzed radar signals within the range profiles behind a Boeing 737 aircraft on February 7, 2021, within the airspace of the Runway Protection Zone (RPZ) at Tianhe Airport, Wuhan, China. The findings revealed that the wake vortex extended up to 6 km from the aircraft, which is 10 km from the radar, displaying distinct stages characterized by scattering patterns and Doppler signatures. Despite the wake vortex exhibiting a scattering power approximately 10 dB lower than that of the aircraft, its Doppler Signal-to-Clutter Ratio (DSCR) values were only 5 dB lower, indicating a notably strong scattering power within a single radar bin. Additionally, certain radar parameters proved inconsistent in the stable detection and tracking of wake vortex, aligning with our earlier concept of cognitive micro-Doppler radar.

## I. INTRODUCTION

Detection of wake vortices trailing an aircraft is feasible; however, prevailing research suggests that the detection range is limited, typically within several kilometers from the radar location. This limitation is often attributed to the perceived insignificantly low radar reflectivity or radar cross-section (RCS) value of wake vortices at longer ranges.

A pivotal study by Gilson, conducted at Lincoln Laboratory in 1992 at Kwajalein atoll, where humidity is enough for wake vortex reported a breakthrough in wake vortex detection [1]. Gilson successfully detected the wake vortex at distances behind the aircraft ranging from about one hundred meters to tens of kilometers, and the wake vortex come from a C-5A, while the wake of the Learjet was too small for detection, even employing exceptionally powerful pulse-Doppler radars generating 2-7 MW of peak power within VHF, UHF, L-, S-, and C-band. Gilson's findings indicated a consistent wake vortex RCS across radar frequencies, with a decrease observed with altitude. The measured RCS values fluctuated within a range of 20 dB, spanning from -70 to -50 dBm², as shown in Table I [1][2]. Yet, they concluded that the wake vortex could be necessity detected with extremely powerful and sensitive instrumentation radars, and the RCS values of wake vortex is too small for most practical applications.

Table I

Kiernan Reentry measurement system radars [1][2]

|  | ALTAIR | | TRADEX | | ALCOR | MMW |
|---|---|---|---|---|---|---|
| Aperture diameter(ft) | 150 | | 84 | | 40 | 45 |
| Radar band | VHF | UHF | L | S | C | Ka |
| Peak power (MW) | 7 | 5 | 2 | 2 | 3 | 0.025 |
| Single pulse sensitivity at 200 km (dBsm) | -64 | -73 | -68 | -58 | -51 | -45 |

Thales presented another notable work employing an X-band radar (BOR-A-550) with peak powers of 20 W / 40 W, complemented by LIDAR confirmation in certain instances. Their research, conducted at Orly Airport in 2006 & 2007 and Paris-CDG Airport in 2008, demonstrated the X-band radar's ability to monitor wake vortex roll-ups across diverse weather conditions, leveraging high-resolution Doppler entropy assessment for sensitive wake vortex detection and localization within short ranges, and measured the wake vortex RCS, approximating 0.01 m² for medium aircraft like the Airbus A320, particularly within short ranges (~2 km) and various weather conditions (wet or dry) [3][4][5]. They concluded that X-band radar can detect and monitor wake vortex of aircraft for different weather conditions, and use the results for suggesting aircraft separation criteria at airports.

Subsequent studies contributed valuable insights. Neece et al. demonstrated wake vortex tracking using a 35 GHz pulse-Doppler radar under low visibility conditions, detecting vortex reflectivity as low as -13 dBz (heavy fog) at a range of 1600 m with 500 W transmitted power [6]. Seliga et al. investigated wake vortex detection during aircraft landings in light rain, utilizing a low-power (100-mW), solid-state W-band (94 GHz) pulse-Doppler radar. Despite its limited range of 1000 m, the radar's high sensitivity (-15 to -10 dBz) and exceptional capabilities in capturing sensitive wake vortex behavior,

This work is supported by Natural Science Foundation of Hubei Providence, Youth Program-2023AFB130.

Jiangkun Gong, Jun Yan, and Deren Li are with State Key State Key Laboratory of Information Engineering in Surveying, Mapping and Remote Sensing, Wuhan University, No. 129 Luoyu Road, Wuhan, China (e-mail: gjk@whu.edu.cn, yanjun_pla@whu.edu.cn, drli@whu.edu.cn. Deyong Kong is with School of Information and Communication Engineering, Hubei Economic University, No.8 Yangqiaohu Road, Wuhan, China (e-mail: kdykong@hbue.edu.cn)

*Corresponding author: Jun Yan (yanjun_pla@whu.edu.cn; +86-027-68778527)

geometry-based radar reflectivity patterns, and Doppler spectra were evident [7].

Despite widespread belief that small wake vortex RCS poses challenges for radar detection, our study challenges this notion. We demonstrate that using an X-band pulse-Doppler radar with a moderate power of 320 W, the detection range of wake vortices trailing civil aircraft can extend beyond 10 km. Our methodology involves aircraft positioning followed by the extraction and analysis of radar signals behind the aircraft. We contend that the radar echoes discerned within the radar resolution behind the aircraft correspond to those originating from wake vortices. We quantify the Signal-to-Noise Ratios (SNRs), Radar Cross-Sections (RCSs), and Doppler velocities associated with these wake vortices.

## II. METHOD AND MATERIALS

### 2.1 Physical characters of wake vortex

The wake vortex characteristics of transport aircraft primarily involve discussions centered on the flow and flight physics of these vortices. Breitsamter provided an overview of these characteristics, dividing the stages of a wake vortex's lifespan into four regions based on the ratio between the distance of the wake vortex from the aircraft and the wingspan. Analyzing wake vortex flow-fields involves various quantities such as mean velocities, mean axial vorticity, circulation, induced velocity, time scales, turbulence intensities, spectral densities, and induced rolling moments, as illustrated in Fig. 1a [8]. The ratio to describe the stage of the wake vortex ($R_{wv}$) is given by:

$$R_{wv} = \frac{x}{b} \quad (1)$$

where $x$ represents the distance behind the aircraft, $b$ denotes the wingspan of the aircraft, and $R_{wv}$ signifies the ratio.

These stages are characterized as follows:

- Near field ($R_{wv} \leq 1$, 'Young'): Characterized by the formation of highly concentrated vortices shed at all surface discontinuities.

- Extended near field ($1 < R_{wv} \leq 10$, 'Mature'): In this stage, the wake roll-up process occurs, leading to the merging of dominant vortices (e.g., those shed at flap edges, wing tips, etc.), gradually resulting in the formation of two counter-rotating vortices.

- Mid field/far field ($10 < R_{wv} \leq 100$, 'Old'): Here, the wake descends in the atmosphere, and linear instabilities start to emerge.

- Decay region ($100 < R_{wv}$, 'Decaying'): In this final stage, fully developed instabilities lead to a strong interaction between the two vortices until their collapse.

Frederic Barbaresco et al. from Thales observed that time-Doppler signatures can be explained by the logarithmic spiral structure of wake vortices, as depicted in Fig. 1b [3]. Roll-ups signify complex air structures formed by the interweaving of air currents from the surrounding environment and higher altitudes. These structures involve the adiabatic transport of fluid within a vortex pair. As each roll-up rotates, the span of reflecting points along each fence expands. This expansion is influenced by the wake vortex's age and tangential speed law, resulting in distinct time/Doppler slope patterns: positive slopes for young vortices, a combination of positive/negative slopes for mature vortices, and negative slopes for old and decaying vortices. In the case of a "young vortex," the wake core appears dense with a noticeable increase in tangential speed corresponding to the radius. Conversely, in an "old vortex," core deterioration due to diffusion is evident, resulting in a decrease in tangential speed as the radius increases.

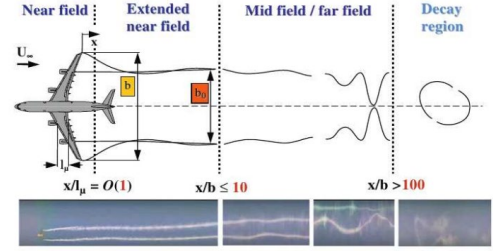

a

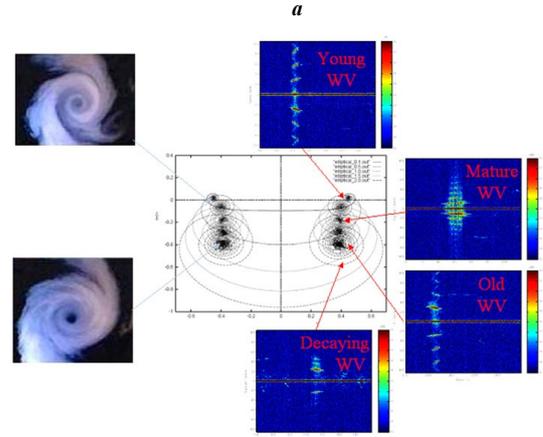

b

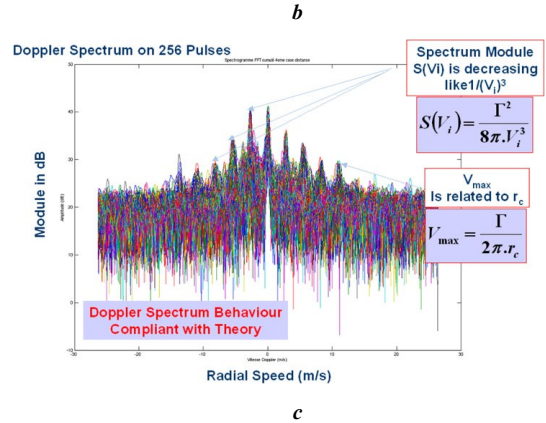

c

*Fig. 1. Stages or ages and Doppler characters of wake vortex,* **(a)** Phases of wake-vortex evolution in cruise flight [8], **(b)** Evolution of roll-up spiral geometry & Doppler spectrum (time/Doppler slopes) versus age [3], **(c)** Doppler FFT for all radar cells in case of wake vortex [3].

### 2.2 Detection method

Based on the real-time status of the Boeing 737 aircraft illustrated in Fig. 2, we utilized radar data to analyze the wake vortex phases following the evolution stages outlined in Fig. 1. Notably, our approach differs from that employed by Frederic Barbaresco et al. from Thales [3]. In their study, they also utilized an X-band radar with parameters detailed in Table II. Their radar system could boast a high velocity resolution of 0.04 m/s, necessitating the use of non-coherent integration technology to observe the same target across multiple instances. This involved accumulating data from eight Coherent Processing Intervals (CPIs), totalling approximately 1.529 seconds. Moreover, it's important to note that the time-scale of Doppler-based images capturing wake vortex phenomena in their study operates at around 10-second intervals. To achieve this, they employed multiple scans, non-coherent integration techniques, and extended radar observation times, typically involving a dozen or more scans.

Unlike Thales' methodology, our approach doesn't depend on non-coherent integration outcomes. Instead, we meticulously analyze radar data within every individual range resolution cell in a single CPI. Consequently, our radar data's time scale is significantly shorter, lasting only 100 milliseconds compared to Thales' tests spanning over 1 second. This divergent methodology enables us to assess wake vortex characteristics within the framework of each radar data segment, avoiding reliance on accumulated data from multiple instances.

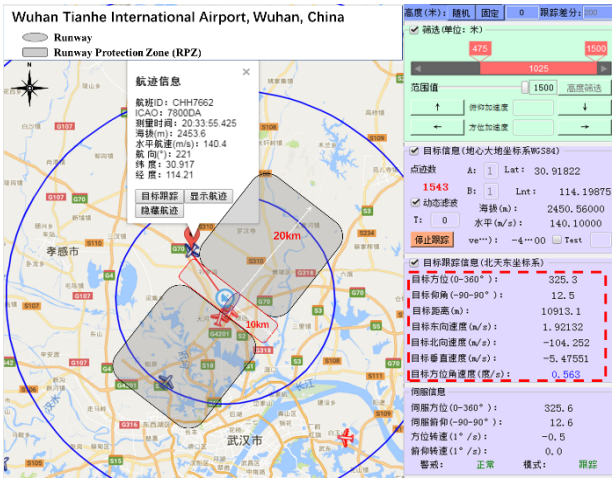

*Fig. 2*. *The aircrafts flying around Tianhe Airport on the radar screen*

TABLE II Comparisons of wake vortex detection radars

| Parameters | Thales | WHU |
| --- | --- | --- |
| Model | BOR-A-550 | WHU-X101 |
| Band | X-band | X-band |
| Peak power | 20 W/ 40 W | 320W |
| Mechanical Scan rate | 8°/s | 0.5°/s |
| PRF | 3.348 kHz | 11.4 kHz |
| CPI | 76.46 ms | 178.1 ms |
| Bandwidth | 3.75 MHz | 5 MHz |
| Range resolution | 40 m | 30 m |
| Sample pulses | 256 | 2048 |
| Velocity resolution | 0.2 m/s or 0.04 m/s | 0.08 m/s |
| | | |
| Detection aircraft | A380 | Boeing 737 |
| Detection range | 2~7[1] km | >10 km |
| Detection height | ~1500 m | ~3452 m |

1. Thales claimed that the detectable radar of the wake vortex is up to 7 km, but mainly 2 km.

Similar to Thales' approach, which extract radar signals of wake vortex using Doppler domain [3], our methodology also involves the extraction of Doppler radar signatures for wake vortex monitoring. However, we enhance this process by integrating automatic target recognition (ATR) technology based on geometric information to identify wake vortices. As the ATR technology is proprietary and beyond the scope of this paper, we provide limited discussion on its details here. The distinct logarithmic spiral structure inherent in wake vortices results in a unique Doppler spectrum, as depicted in Fig. 1c. This spectrum exhibits multiple Doppler peaks, corresponding to various layers of the wake vortex. These peaks display a discernible pattern in their distribution, both in magnitude and intervals between adjacent peaks. In our earlier publications [9], we introduced the concept of the radar Doppler signal-to-clutter ratio (DSCR) detector, which we reference here as:

$$DSCR(D) = 10 log 10 [\frac{F(D)}{\frac{\sum_1^N F(K)}{N}}] \qquad (2)$$

Where:

- $F(K)$ denotes the amplitude of frequencies in the spectrum of the current radar bin.

- $K$ represents the Doppler frequency in the spectrum.

- $N$ stands for the length of the spectrum.

- $D$ could refer to the body Doppler, micro-Doppler, or any Doppler within the spectrum. Here, we select the Doppler with the highest magnitude as $D$ after eliminating clutter around zero hertz.

*2.3 Test environment*

As shown in Table II, the radar system utilized in this study is an X-band (10.1 GHz) air surveillance radar equipped with an Active Electronically Scanned Array (AESA) antenna. Distinguished by its software-defined architecture, this radar leverages phased array technology, providing it with cognitive capabilities designed for aircraft surveillance and counter-drone applications. Through its software-defined approach, the radar system exhibits adaptability in key parameters such as CPI, PRF, and bandwidth, allowing customization tailored to specific mission requirements. In this paper, we mainly utilized data obtained at a Coherent Processing Interval (CPI) of 178.1 ms and a Pulse Repetition Frequency (PRF) of 11.4 kHz. The radar system achieves a range resolution of 30 meters, enabling precise discrimination and analysis of radar echoes within specific distance intervals.

The original data collection site is situated in the vicinity of Tianhe Airport, Wuhan, China. Illustrated in Fig. 2 and detailed in Table III, the radar station is positioned outside the airport perimeter, monitoring aircraft within the approach and departure airspace. For this study, our focus is on a cooperative target, namely a civil aircraft: the Boeing 737-800. This aircraft serves as a short-to-medium-haul workhorse, equipped with two jet engines, boasting a wingspan of 34.32 meters and a length from head to tail measuring 38.47 meters. The ADS-B response of this aircraft is visualized in Fig. 2. Its ICAO ID is 7800DA, and it operates under the flight number CHH7662. The aircraft's flight data, recorded at 20:33:55.425 on February 7, 2021, indicates a radar-detected altitude of 10.901 kilometers and a location at 2453.6 meters Above Sea Level (ASL), positioned at a longitude of 114.19875 and a latitude of 30.91822. The aircraft's flying velocity is approximately 140.1 meters per second.

TABLE III

Test conditions

| Date & Time | 2021-2-7, 7:00 PM-8:00 PM |
|---|---|
| Location | 114.19875 W, 30.91822 N |
| Temperature | 50 |
| Humidity | 100 |
| Wind | NNW |
| Wind speed | 4 mph (1.78 m/s) |
| Wind gust | 0 mph |
| Pressure | 30.11 in |
| Precipitation | 0 in |
| Condition | Light rain |

III. RESULTS

The wake vortex initiates from the radar bin where the Boeing 737 aircraft is positioned. Fig. 3 depicts detection results obtained using range-Doppler images, featuring 512 range windows. Each range window extends over 15.3 km with a precise range resolution of 30 meters. The data collection parameters include a CPI of 178.1 ms and a PRF of 11.4 kHz. The aircraft is situated within the 286th range cell. In the illustrated range-Doppler images, the green section represents the range-Doppler view, the upper segment signifies the sampled data along the range (range-profile), and the left portion denotes the sampled data along the Doppler (Doppler-profile). Within the range-Doppler images, the wake vortex's trail behind the Boeing 737 aircraft is discernible, spanning a length of up to 6 km.

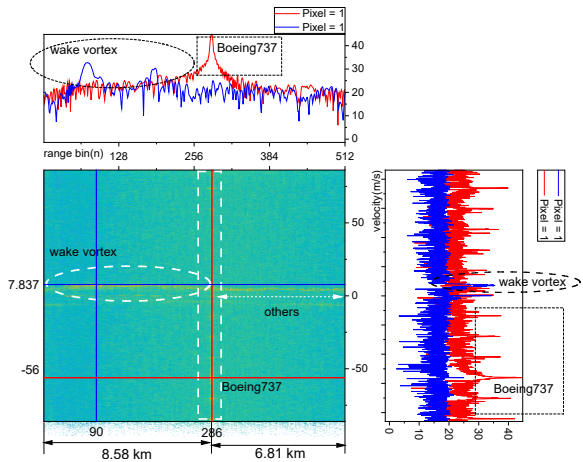

*Fig. 3. The range-Doppler image of radar data*

Notably, both in the time domain (range-profile) and frequency domain (Doppler-profile), the magnitude values associated with the aircraft exceed those of the wake vortex. Specifically, within the range-profile image, the wake vortex spanning the 1st to the 285th range cells exhibit fluctuating amplitudes, trailing behind the aircraft positioned in the 286th range cell. All wake vortex amplitudes measured within this range fall below those of the aircraft in the 286th range cell. The aircraft's relative Doppler velocity is approximately -56.0 m/s, indicating movement away from the radar location. This Doppler velocity is ambiguous. Additionally, conspicuous jet engine modulation (JEM) Dopplers are observed within the aircraft's spectrum in the Doppler-profile image. In contrast, the wake vortex velocity registers around 7.8 m/s, significantly lower than that of the aircraft, but much quicker than the wind' speed around 1.78 m/s (Table III). It's essential to note that the area preceding the aircraft, spanning the 287th to the 512th range cells, is not attributed to the wake vortex of this aircraft. Subsequent sections will elucidate the reasons behind this distinction.

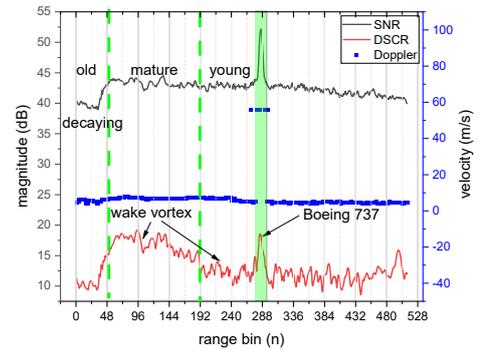

*Fig. 4. Comparison of SNR values in time domain and DSCR values in frequency domain of aircrafts and wake vortex in one range-window*

The findings reveal that the magnitude of the wake vortex trailing the aircraft can be notably weak, yet its Doppler magnitudes considerably surpass those of the aircraft. Fig. 4 presents a comparison among Signal-to-Noise Ratio (SNR) values, Doppler Signal-to-Clutter Ratio (DSCR) values, and Doppler velocities across each radar cell within the range window illustrated in Fig. 2. The aircraft and the near-field wake vortex, as depicted in Fig. 1, are delineated by the green shadow, while dotted green lines demarcate various wake vortex stages extending from the 1st to the 286th range cell.

TABLE IV

Comparisons of detection values

| Value | Length (km) | SNR (dB) | DSCR (dB) | Velocity (m/s) |
|---|---|---|---|---|
| Old/Decaying | 1.44 | 40.18 | 10.99 | 5.53 |
| Mature | 4.3 | 43.29 | 16.72 | 6.93 |
| Yong | 2.5 | 42.84 | 12.33 | 6.89 |
| Aircraft | 0.06 | 52.35 | 17.84 | 56.06 |
| others | 6.4 | 41.85 | 11.77 | 4.58 |

The mean values of different wake vortex ages and the aircraft's metrics are calculated and summarized in Table IV. The range separation of different stage of the wake vortex is based on Fig. 1a. However, if we overlook the varying ages of the wake vortex and compute the mean values across its entire lifespan, the wake vortex demonstrates an average SNR value of about 42.19 dB, approximately 10 dB lower than the aircraft's 52.35 dB. Similarly, the average DSCR value for the wake vortex measures around 13.34 dB, indicating a discrepancy of approximately 4.4 dB compared to the aircraft's 17.84 dB. It's crucial to note that this generalization disregards the diverse stages of the wake vortex, which exhibit slight variations. For instance, the DSCR of a mature wake vortex is comparable to the aircraft's DSCR, signaling robust scattering characteristics.

Additionally, the radar signals in the range cells spanning from the 287th to the 512th in Fig. 3 & 4, labeled as 'others' before the aircraft in Table IV, likely do not represent the wake vortex of this particular aircraft but could potentially originate from other aircraft. Given the monitoring location within the Runway Protection Zone (RPZ) of Tianhe airport (Fig. 2), multiple aircraft continuously traverse this airspace, generating distinct wake vortices. Therefore, these radar signals potentially correspond to wake vortices produced by other aircraft. Firstly, the SNR values of these signals labeled as 'others' bear resemblance to wake vortex values, suggesting the presence of 'ghost' objects resembling wake vortices above the aircraft. Please note that these SNR values represent the logarithmic transformation of the raw received power. Secondly, the DSCR values of these signals labeled as 'others' are lower than the wake vortex values, coupled with lower velocity values, indicating the possibility that these signals may represent older wake vortex trails or those generated by aircraft that flew through the area at a much earlier time.

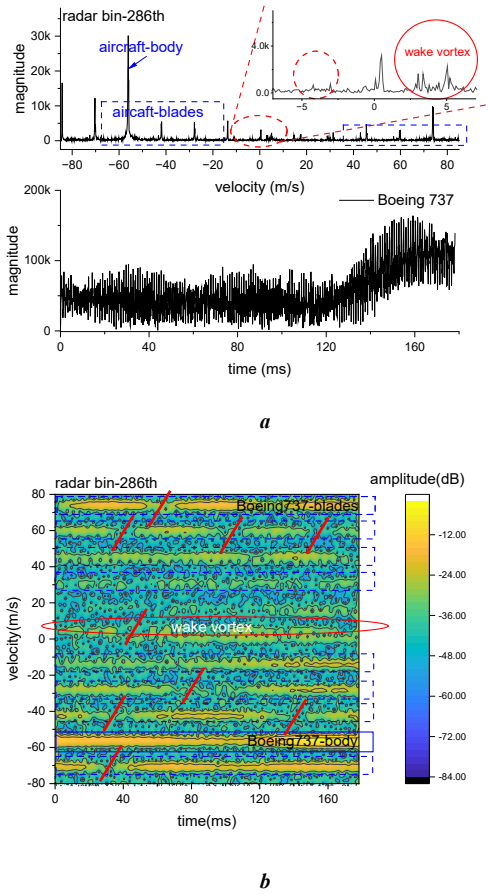

*Fig. 5.* Radar data of the aircraft in 286th radar bin, **(a)** signals in time and frequency domain, **(b)** micro-Doppler images.

Radar signatures of the wake vortex and the aircraft are different. In Fig. 5, the micro-Doppler signatures in both the time-frequency domain and the micro-Doppler image corroborate the detection of the Boeing 737 aircraft within the 286th range bin. Notably, the body Doppler in Fig. 5a registers at -56.0 m/s, contrasting the 140 m/s velocity recorded by ADS-B, attributable to the inherent ambiguity in Doppler velocity measurements. Within the spectrum, alongside the primary Doppler, numerous micro-Dopplers are evident. These shares patterned magnitudes with approximate intervals, signifying the presence of Jet Engine Modulation (JEM) Dopplers modulated by the aircraft's jet engine [10][11][12]. Notably, the predominant JEM Dopplers, exhibiting higher magnitudes, pertain to the first stage blades of the engine, while the smaller ones correspond to the second stage blades, as detailed in Table V. Furthermore, among these Dopplers, those registering at 3.9 m/s, 5.0 m/s, 5.2 m/s, etc. are indicative of newly formed wake vortex. These nascent wake vortices manifest as small in size and exhibit slower Doppler velocities. The Doppler groups evident in the spectrum illustrate the presence of multiple eddy layers within the wake vortex. The micro-Doppler spectrogram delineates the temporal distribution characteristics of these Dopplers. It's worth noting that while most aircraft's body Doppler belt span the entire 178.1 ms within one CPI, the Doppler belt of the wake vortex appears intermittent in its existence.

TABLE V
Components of different Dopplers in the spectrum

| Sources | | Dopplers (m/s) |
|---|---|---|
| Body | | -56.0 |
| JEM | First | -84.2, -72.0, -43.2, -27.8, -13.7, 17.5, 28.8, 31.6, 45.4, 59.9, 73.9, etc. |
| | Second | -41.8, -10.5, 0.5, 14.5, 43.0, 56.9, 83.4, etc. |
| Wake vortex | | -4.2, -3.0, 3.0, 5.0, 5.2 |

Wake vortex in different stages exhibits distinct radar signatures compared to those of the aircraft. In Fig. 6, Doppler spectrums and micro-Doppler spectrograms for various ages of wake vortex, spanning radar bins from the 1st to the 285th behind the aircraft, are illustrated. The vertical and horizontal axis values remain consistent across all images, with circles marked at identical positions in each image. Fig. 6a showcases three distinct Doppler spectrums representing wake vortex at different ages—ranging from the young to mature to old stages, correlating with the stages identified in Fig. 4. The young wake vortex, situated in the 283th radar bin, bears resemblance to the spectrum in Fig. 5a but features more wake-vortex Dopplers around -5.0 m/s and 5.0 m/s at a magnitude level of 1 k. Positioned approximately 150 meters away from the newborn wake vortex and the aircraft in Fig. 5a, the young wake vortex continues its development towards maturity. The mature wake vortex, situated in the 90th radar bin about 5.8 kilometers away from the newborn wake vortex and the aircraft, exhibits Dopplers with stronger magnitudes at the 3 k level and velocities around 7 m/s—both higher and faster than those observed in the young stage. Furthermore, the negative Dopplers in the mature wake vortex appear to diminish, indicating robust scattering power. In the final stage, the old wake vortex exhibits weaker magnitudes around 500. However, both negative and positive Dopplers are detected, with the positive Dopplers displaying a wider spread compared to the mature and young stages.

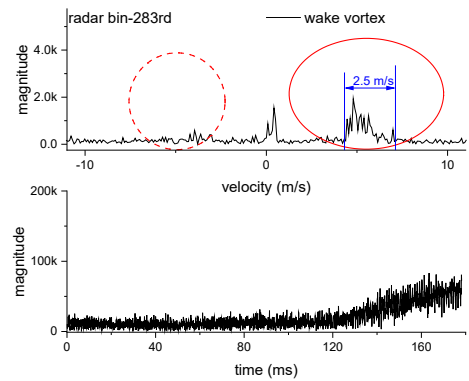

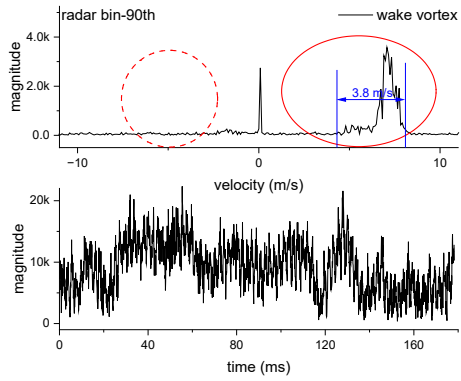
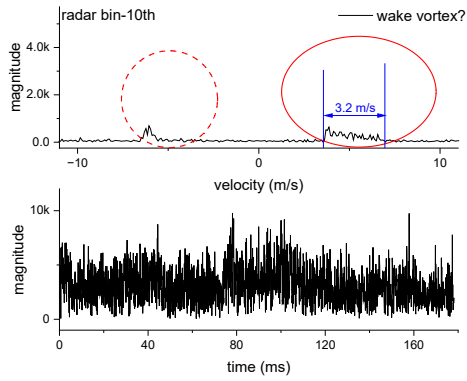

*a*

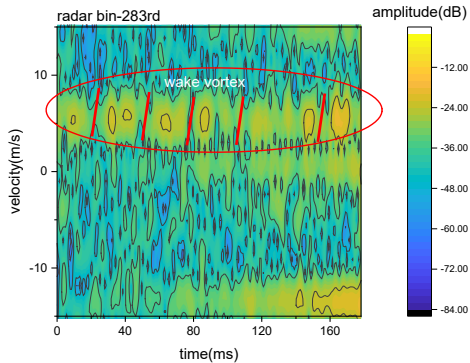
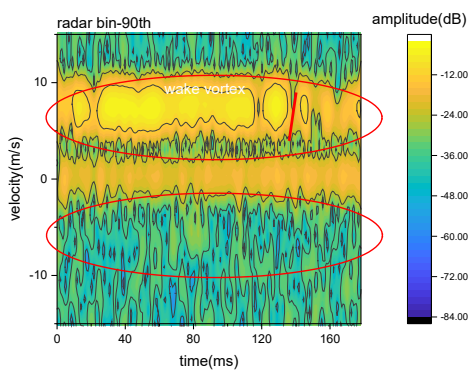

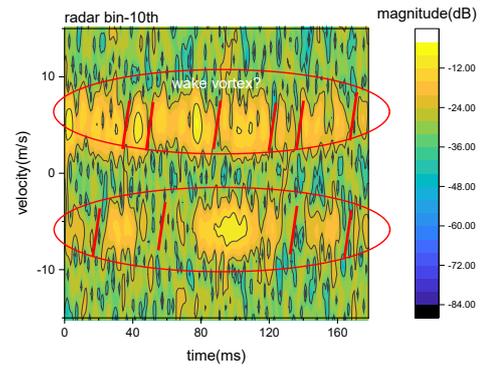

*b*

*Fig. 6. Typical radar data of the wake vortex in different ages*, (a) signals in time and frequency domain, (b) micro-Doppler images.

The micro-Doppler spectrograms portray a vivid evolution pattern of the wake vortex (Fig. 6b). The young stage displays intermittent speckles, indicating the growing multilayers, where no single layer remains consistently dominant. As the wake vortex matures, the strongest layer reflects radar waves consistently over most radar dwell times, implying stability. Conversely, in the old stage, as the wake vortex begins to decay, it becomes unstable, featuring numerous intermittent speckles in the belts. Unlike the Doppler spectrums, the spectrograms are less quantifiable mathematically but provide a more intuitive human-readable representation, depicting changing Doppler patterns from different layers of the wake vortex over time.

The radar signals preceding the aircraft likely originate from wake vortex created by other aircraft. Fig. 7a compares Doppler spectrums from the 481st and 90th radar cells, both situated approximately 5.8 kilometers from the aircraft. The 481st radar cell displays Doppler groups similar to those observed in the wake vortex of the 90th radar cell, yet with weaker magnitudes (1k < 4k) and slightly slower velocities (4.0 m/s < 7.0 m/s). Notably, while the 90th radar cell features a -56.0 m/s Doppler marking the aircraft, such a signal is absent in the 481st radar cell, affirming that it doesn't originate from the aircraft. The Doppler and micro-Doppler spectrograms in Fig. 7b suggest that this signal may indeed correspond to wake vortex, potentially emanating from passing aircraft, considering the airspace's location within the Runway Protection Zone (RPZ) of Tianhe airport.

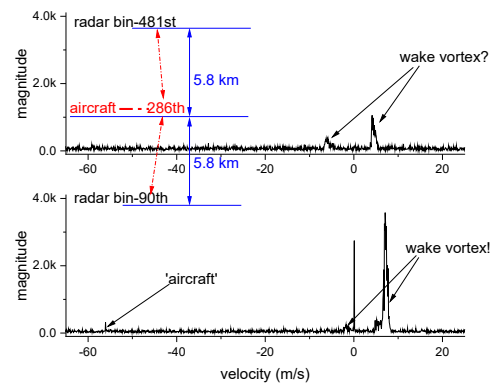

*a*

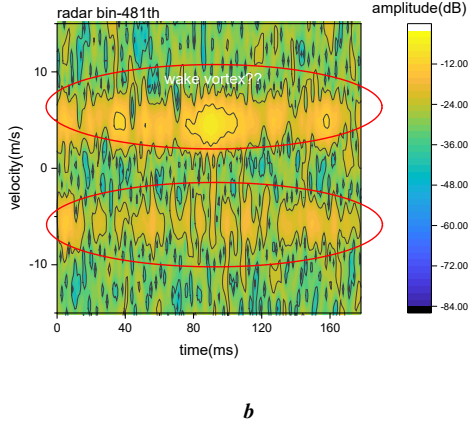

*b*

*Fig. 7. Typical radar data of in other radar bins ahead the aircraft, (a) comparison it with signals in 90th radar bin, (b) micro-Doppler image of the signals in 481th radar bin.*

The aircraft tracking results confirm the stable radar detection of the wake vortex. Illustrated in Fig. 8, the mean tracking update occurs every 0.5 seconds. Given the aircraft's speed of approximately 140 m/s, it traverses a distance of only 70 meters, which is roughly twice the length of a single radar range resolution. Consequently, the aircraft's locations predominantly center around the 286th range bin, close to the midpoint of the 512-range bin window.

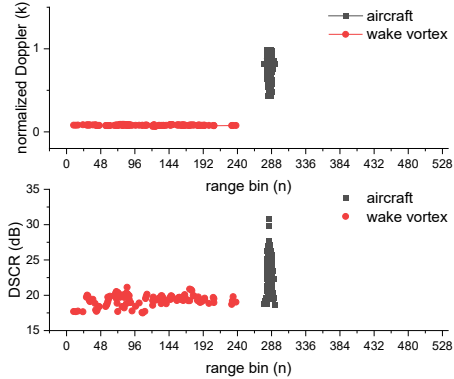

*Fig. 8. Tracking results of the aircraft and the following wake vortex*

As previously mentioned, our ATR module effectively discerns radar signals from the wake vortex and the aircraft based on specific signatures. Initial statistical data reveals that the wake vortex consistently trails the aircraft, primarily spanning range bins 1 to 285, while signals ahead of the aircraft do not exhibit characteristics indicative of the wake vortex. Furthermore, distinctions between the wake vortex and the aircraft include lower & homogeneous DSCR values and slower but more consistent Doppler velocities in the wake vortex compared to the aircraft. Despite these differences, the tracking data supports the notion that detecting wake vortex behind an aircraft is achievable using an X-band radar with moderate transmitted power, achieving a detection range of approximately 10 km and tracking over a range scale exceeding 6 km.

## IV. DISCUSSIONS

It's remarkable that our radar can detect the wake vortex of a medium-sized aircraft at a range of over 10 km. Primarily, our radar exhibits comparable wake vortex detection range capabilities to Thales' radar, as illustrated in Table II. Theoretically, according to the radar equation, the radar detection range of a specified target is proportional to the fourth root of its RCS. According to the radar equation, the detection range, $R$, is calculated by [10],

$$R = \sqrt[4]{\frac{P_t G A_e \sigma}{(4\pi)^2 k T_0 B_n F_n C_{SNR}}} \quad (3)$$

where $P_t$ is the transmit power; $A_e$ is the effective area of the receiving antenna; $B_n$ is the noise bandwidth of the receiver; $\sigma$ is the RCS of the target; $T_0$ is the standard temperature; $F_n$ is the noise figure; and $C_{SNR}$ is the signal-to-noise ratio (SNR) value. Therefore, with a rough estimation based on our radar's peak power of 320 W, approximately 16 times that of the Thales' radar, the theoretical detection range for an object at 7 km would be around 14 km. Consequently, our radar's wake vortex detection capability seems to align closely with the capacity of Thales' radar.

Secondly, we examine radar signals within radar bins situated behind the wake vortex, considering the aircraft's location as a reference point. Given the wake vortex's distinctive Doppler spectrums (refer to Fig. 1c), it becomes detectable once the radar detector, based on Doppler analysis, is engaged. Employing the Doppler Spectrum Covariance Ratio (DSCR) detector enables us to isolate the Doppler signals emitted by the wake vortex for subsequent analysis. As depicted in Fig. 4 and Table IV, the Signal-to-Noise Ratio (SNR) value associated with the wake vortex registers > 10 dB lower than that of the aircraft. However, the DSCR value is approximately 5 dB lower. This discrepancy implies that the scattering power of the wake vortex is more pronounced in the Doppler domain compared to the time domain. Essentially, although the wake vortex exhibits a small Radar Cross Section (RCS), its robust Doppler signature can extend the radar detection range, especially when detected using Doppler-related detectors.

Furthermore, due to the wake vortex's considerable size, RCS becomes an inadequate parameter for depicting its scattering power. Mentioning the RCS of the wake vortex without considering its spatial scale would be meaningless and misleading. Hence, parameters like radar reflectivity [13] would be more suitable as they are designed for describing objects (e.g., birds, clouds, chaff, etc.) with a larger scale.

Not all radar parameters ensure stable detection and tracking of wake vortex. Changes in Coherent Processing Interval (CPI) and Pulse Repetition Frequency (PRF) values significantly impact wake vortex detection. Fig. 8 compares two scenarios utilizing the same radar system but varying CPI and PRF settings. The mean tracking update is about 0.5 s. Specifically, the track frame shifts from the 15th to the 16th tracking frame within 1s while altering from (CPI = 178.1 ms, PRF = 11.4 kHz) to (CPI = 233.4 ms, PRF = 8.7 kHz). Radar bins located approximately 30 meters away from the aircraft are extracted for analysis, encompassing radar signals from both the aircraft and the newly formed wake vortex.

Upon analysis, it's evident that the radar configuration registered in track-15th (CPI = 178.1 ms, PRF = 11.4 kHz) successfully detects wake vortex velocities around 5-7 m/s. However, the alternate configuration noted in track-16th (CPI = 233.4 ms, PRF = 8.7 kHz) displays considerably weakened wake vortex detection. This attenuation in detection capability is potentially due to the micro-Doppler signals reflected by jet blades of the aircraft, which might be suppressing the wake vortex signals within the spectra.

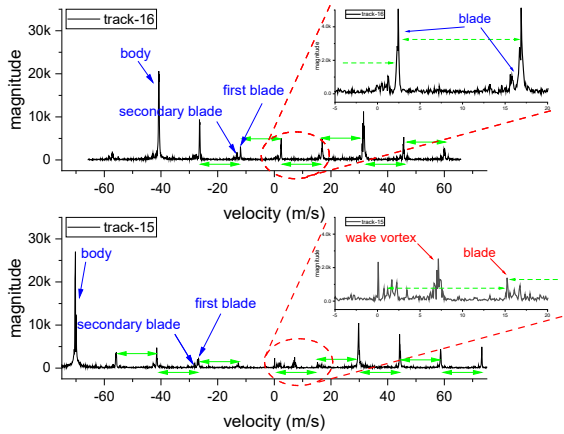

**Fig. 9.** *Typical radar data of the aircraft and wake vortex using different radar waveforms.*

Our previous endeavors introduced the concept of cognitive micro-Doppler radar, showcasing its practical implications through specific scenarios. Examining the radar's detection of micro-Doppler signals, particularly JEM Dopplers modulated by jet-engine blades, reveals diverse outcomes with different radar configurations. Table VI presents the detection results of JEM Dopplers using two distinct radar setups.

The green lines in Fig. 9 demarcate intervals between adjacent first blade JEM Dopplers, consistently around 14.3 m/s. Notably, the radar configuration recorded in track-16th (CPI = 233.4 ms, PRF = 8.7 kHz) efficiently and clearly identifies both first and secondary blade micro-Dopplers, averaging approximately 2.9 m/s between them. Conversely, the radar setup in track-15th (CPI = 178.1 ms, PRF = 11.4 kHz) easily discerns micro-Dopplers of first stage blades with a mean interval of 14.4 m/s but struggles with detecting that of secondary stage blades. Intriguingly, while track-15th demonstrates poor performance in detecting aircraft JEM Dopplers, it exhibits better wake vortex detection compared to track-16th.

TABLE VI

Components of aircraft Dopplers using different radar waveforms

| Sources | | Dopplers (m/s) |
|---|---|---|
| Track-16 | Body | -40.7 |
| | First | -26.2, -11.9, 2.4, 16.9, 31.4, 45.6, 60.0, etc. |
| | Secondary | -27.6, -13.2, 1.2, 15.8, 30.2, 44.5, 58.9, etc. |
| Track-15 | Body | -70.2 |
| | First | -55.9, -41.5, -27.1, -12.7, 0.0, 15.2, 29.7, 44.1, 58.7, 73.2, etc. |
| | Secondary | -42.8, -28.3, etc. |

This phenomenon is elucidated by the target's scattering mechanism, which involves alterations in its geometry or micro-motions, demonstrating practical applications in the development of cognitive micro-Doppler technology. The micro-motions originate from the aircraft's jet engine and the evolving layers within the wake vortex. When a radar incident wave, comprising numerous pulses, interacts with the object, more radar signals illuminate the primary body than the micro-areas, impacting the behavior of micro-Dopplers across different radar waveforms. The CPI and PRF in pulse-Doppler radar significantly influence signal intensity. Careful selection of radar waves can enhance detection and identification of specific targets, exemplified by the distinct performance of the two radar configurations.

The concept of cognitive micro-Doppler radar underscores the idea of tailoring radar waveforms settings for improved detection and identification. Future endeavors will focus on ongoing analysis of wake vortex radar data. Presently, Doppler signatures indicate the presence of wake vortex but fall short for radar recognition. Modeling Doppler distributions associated with evolving wake vortex eddies might reveal additional signatures beneficial for an Automatic Target Recognition (ATR) algorithm. Expanding data collection to encompass diverse weather conditions and different times of day aims to evaluate wake vortex radar detection performance comprehensively. Additionally, forthcoming research will explore the application of cognitive micro-Doppler radar in understanding wake vortex characteristics more deeply.

## V. CONCLUSION

The wake vortex's detection range extends beyond 10 km from the radar location. This study presents the detection and tracking findings of a wake vortex trailing a medium-sized aircraft, specifically the Boeing 737, using an X-band pulse-Doppler radar equipped with modest peak power (320 W) and a novel radar detector using DSCR values, within the airspace of the Runway Protection Zone (RPZ) at Tianhe Airport, Wuhan, China. The radar effectively spans a range exceeding 10 km, capturing a wake vortex stretching up to 6 km behind the aircraft. Doppler signatures and micro-Doppler signals facilitate the classification of different wake vortex stages. Preliminary analysis suggests that the characteristics of radar waveforms influence the wake vortex detection and tracking performance. This insight highlights the potential for employing cognitive micro-Doppler radar to delve deeper into understanding the formation and evolution of wake vortices trailing behind aircraft.